\documentstyle[twocolumn,aps,epsfig]{revtex}

\begin{document}
\title{Black Hole Production in Large Extra Dimensions at the Tevatron:\\
Possibility for a First Glimpse on TeV Scale Gravity}

\author{Marcus Bleicher\ddag, Stefan Hofmann\dag,  
Sabine Hossenfelder\dag, Horst St\"ocker\dag}

\address{\ddag\ SUBATECH,
Laboratoire de Physique Subatomique et des
Technologies Associ\'ees \\
University of Nantes - IN2P3/CNRS - Ecole des Mines de Nantes \\
4 rue Alfred Kastler, F-44072 Nantes, Cedex 03, France}

\address{\dag\ Institut f\"ur Theoretische Physik\\
J. W. Goethe Universit\"at\\
60054 Frankfurt am Main, Germany}

\maketitle

\noindent
\begin{abstract}
The production of black holes in large extra dimensions
is studied for Tevatron energies.
We find that black holes may have already been created in small
abundance in $\overline p p$ collisions at $\sqrt s=1.8$~TeV.
For the next Tevatron run ($\sqrt s=2.0$~TeV) large production rates for
black holes are predicted. 
\vspace{1.cm}
\end{abstract}

Recently the possibility of black hole production in large extra dimension 
(LXD) at LHC and from cosmic rays has received great attention 
\cite{Banks:1999gd,gid,dim,Hossenfelder:2001dn,Hofmann:2001pz,Feng:2001ib,Emparan:2001kf,Anchordoqui:2001cg,Ringwald:2001vk,Uehara:2001yk,Landsberg:2001sj}.
In these LXD scenarios\cite{add} the Standard Model of particle physics
is localized on a three dimensional brane in a higher
dimensional space. One scenario for realizing TeV scale gravity is a
brane world in which the Standard Model particles
including gauge degrees of freedom reside on a 
3-brane within a flat compact space of volume
$V_d$, where $d$ is the number of compactified spatial
extra dimensions with radius $L$.
Gravity propagates in both the compact and non-compact
dimensions. The fundamental $D=4+d$ dimensional scale $M_f$ is then
connected to the $4$ dimensional Planck scale $M_{\rm Pl}$ via \cite{add}
\begin{equation}
M_{\rm Pl}^2 = M_f^{2+d} V_d \quad.
\end{equation} 
This raises the exciting possibility
that the fundamental Planck scale $M_f$ can be as low as $m_W$.
As a consequence, future high energy colliders like LHC, CLIC or TESLA
could probe the scale of quantum gravity with its exciting new 
phenomena, namely the production of black holes in high 
energetic interactions (for LHC energies see \cite{dim}).
However, the experimental bounds on the Planck mass in the 
presently discussed scenarios from absence of missing
energy signatures is much lower: $M_f\ge 0.8$~TeV for two to six extra
dimensions \cite{Giudice:1998ck,Mirabelli:1998rt,Peskin:2000ti}.
Astrophysical bounds seem to be more stringent and require
fundamental scales of order 10-100~TeV for two extra dimensions, larger
limits might also be advocated by direct measurements of Newtons law in the
sub-millimeter region or proton stability. However, those limits can be
overcome by increasing the number of extra dimension or invoking additional 
theoretical assumptions (for a discussion see e.g. \cite{Rubakov:2001kp}).
Thus, we will consider only the well known collider bounds on $M_f$ as 
limits for our present investigation.

In this letter we investigate whether a first glimpse on a 
(sub-)TeV scale gravity associated with black hole production might
already be observable at the Tevatron. As discussed elsewhere 
(see e.g. \cite{my,dim,Hossenfelder:2001dn})
the horizon radius of a black hole is given by
\begin{equation}
R_H^{1+d}=
\frac{8 \Gamma\left(\frac{3+d}{2}\right)}
{(2+d)\pi^{\frac{1+d}{2}}} 
\left(\frac{1}{M_f}\right)^{1+d} \; \frac{M}{M_f}
\end{equation}
with $M$ denoting the black hole mass.
The production rate  black holes is 
classically given by \cite{Banks:1999gd,Thorne:1972ji,Giddings:2001ih}\footnote{%
Note that the given  
classical estimate of the black hole production 
cross section is still under debate\cite{Voloshin:2001fe,Giddings:2001ih}.
However, for the present calculations the maximal suppression in the cross
section is by a factor $10^{-1}$\cite{Rizzo:2001dk}, which 
does not invalidate the present arguments.
} 
$\sigma(M)\approx \pi R_H^2$. 
\begin{figure}[h]
\vspace*{-.5cm}
\centerline{\psfig{figure=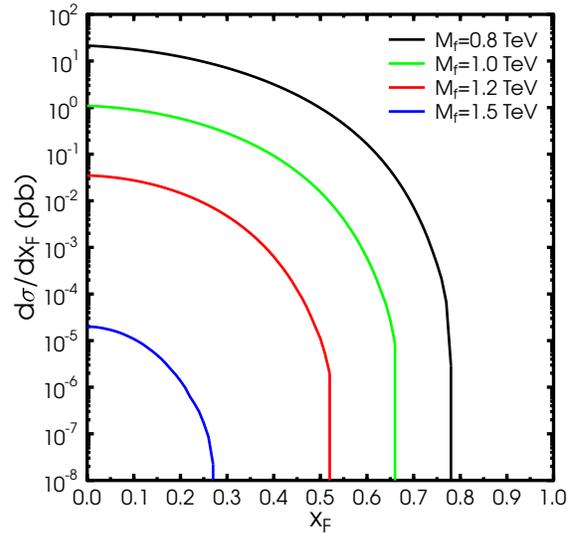,width=3.5in}}
\vskip 2mm
\caption{Feynman $x$ distribution of black holes  
with $M\ge M_f$~TeV produced in pp interactions at the Tevatron ($\sqrt s=1.8$~TeV) 
with four compactified spatial extra dimensions 
and different fundamental scales $M_f$=0.8~TeV, 1.0~TeV, 1.2~TeV and
1.5~TeV (from top to bottom). 
\label{dndy}}
\end{figure}

Since a theory of quantum gravity is still not known, the formation 
of black holes with masses of the order of the fundamental scale 
can  not be justified from first principles. However, in the 
following we assume the applicability of classical gravity. 
As a consequence our predictions are order of magnitude 
estimates for the most optimistic physical scenario.
We will also  neglect complications due to the finite angular 
momentum of the black hole 
and assume non-spinning black holes (roughly a factor two uncertainty).
The influence of finite angular momentum on the
formation and evaporation process of black holes is studied
in \cite{Anchordoqui:2001cg,Hossi2002}.
With these assumptions we ask whether black holes might already have been created
at the Tevatron in Run I and predict their formation cross section for Run II.

By standard methods, the Feynman $x_F$ distribution
of black holes for masses from 
$M\in[M_f,\sqrt s=1.8~{\rm TeV}]$ 
is given by
\begin{eqnarray}
\frac{d\sigma}{d x_F} &=& \sum\limits_{p_1,p_2} 
\int\limits_{M_f}^{\sqrt s} dy \\ 
&&\frac{2 y}{x_1 s} f_1(x_1, Q^2) f_2(x_2, Q^2) 
\sigma(y,d)\; ,
\end{eqnarray}
with $x_F=x_2-x_1$ and the restriction 
$x_1 x_2 s=M^2$.
Where the CTEQ4 \cite{lai1} parton distribution functions $f_1$, $f_2$ 
with $Q^2=M^2$ are used. 
All kinematic combinations of partons from 
projectile $p_1$ and target $p_2$ are summed over. 
\begin{figure}[h]
\vspace*{-.8cm}
\centerline{\psfig{figure=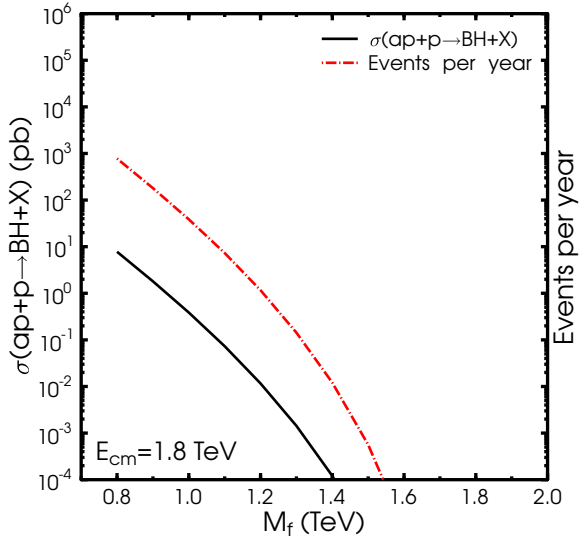,width=3.5in}}
\vskip 2mm
\caption{Black hole production cross section (full line) 
and black hole yield per year (dashed line) at Tevatron as a function
of the fundamental scale $M_f$ for $d=4$ extra dimensions, $\sqrt s=1.8$~TeV, 
${\cal L}=100$~pb$^{-1}$.
\label{sigma1.8}}
\end{figure}

Fig. \ref{dndy} depicts the momentum distribution of produced 
black holes in pp interactions at $\sqrt s = 1.8$~TeV.
Most black holes are of lowest mass ($M_{\rm BH}\approx M_f$) 
and are formed in scattering processes of valence quarks. 
Here we show the result for $d=4$ extra spatial extra dimensions.
A strong dependence on the fundamental gravity scale $M_f$ is 
observed. For the lowest possible $M_f \approx 800$~GeV, significant 
black hole production in $\overline p p$ at $\sqrt s=1.8$~TeV is
predicted.  Higher fundamental scales suppress the production of black
holes at the Tevatron strongly.

Fig. \ref{sigma1.8} shows the production cross section for black holes
as a function of the fundamental scale $M_f$ (full line).
Using an integrated luminosity of $100~{\rm pb}^{-1}$ per year, the 
dashed line gives the expected abundance of black holes produced at Tevatron
per year.
For most optimistic values of $M_f \approx 0.8-1.2$~TeV signals of black hole creation 
might have be observable in past or present day experimental data.
\begin{figure}[h]
\vspace*{-.8cm}
\centerline{\psfig{figure=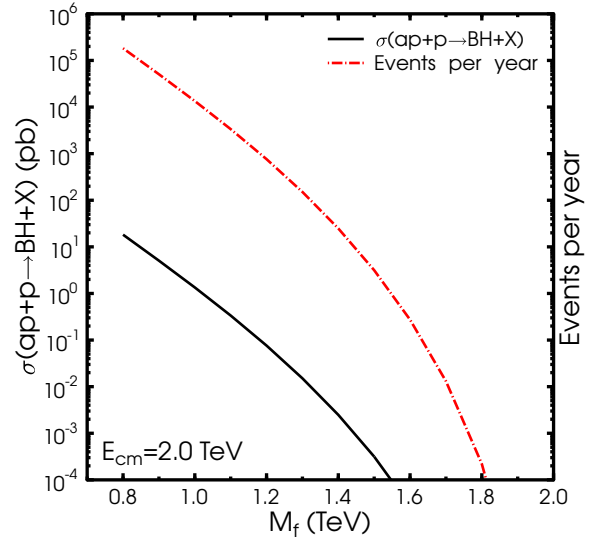,width=3.5in}}
\vskip 2mm
\caption{Black hole production cross section (full line) 
and black hole yield per year (dashed line) at Tevatron as a function
of the fundamental scale $M_f$ for $d=4$ extra dimensions, $\sqrt s=2.0$~TeV, 
${\cal L}=10$~fb$^{-1}$.
\label{sigma2.0}}
\end{figure}

With the update of the Tevatron for Run II higher luminosities at an increased
center of mass energy ($p\overline p$ at $\sqrt s=2.0$~TeV) are available.
Prediction of black hole creation cross sections 
for these runs are shown in Fig. \ref{sigma2.0} as a function of the 
fundamental scale $M_f$ (full line).
Here, an integrated luminosity of $10~{\rm fb}^{-1}$ per year is expected, which
translates into a huge abundance of black holes produced at Tevatron
per year (dashed line).
If the fundamental scale of gravity is $M_f$ is below\footnote{%
The actual identification of a black hole is a difficult task: detailed studies 
about the signal characteristics, backgrounds and experimental cuts are necessary.
For the lightest black holes discussed here this is even theoretically difficult, 
without further knowledge of the quantum theory of gravity. It is clearly out of the
scope of this letter and will be neglected here.}
 $\approx 1.5$~TeV, first signals of black hole creation might be observed at Tevatron before 
the start of LHC.

Note that it has been argued whether it is possible
to observe the emission spectrum of a black hole directly,
since most of the energy maybe radiated in
Kaluza-Klein modes. However, from the higher dimensional 
perspective this seems to be incorrect and most of the energy
goes into modes on the brane\cite{ehm}. 
The life time of the black holes is predicted to
be $\approx 10$~fm/c \cite{Hossenfelder:2001dn}.
It can be observed by the emission of multiple jets
with energies of $\approx 100-150$~GeV  or might
even produce new kinds of elementary particles \cite{Landsberg:2001sj}.
 
In conclusion, the production cross section and the
momentum distribution of black holes in space times with large and compact 
extra dimensions has been discussed. 
It has been demonstrated that for most optimistic values for the mass 
scale of quantum gravity, a first glimpse on black hole production might 
be possible from todays Tevatron ($\sqrt s =1.8$~TeV) data.
For the Run II ($\sqrt s =2.0$~TeV) the calculation predicts the production 
of several thousand black holes per year, if the fundamental 
scale $M_f$ is not too high. This might make the study of TeV scale gravity 
experimentally accessible long before the start of LHC.

This work was supported in parts by BMBF, DFG, and GSI.

\end{document}